\documentclass{iaus}
\usepackage{graphicx}
   \usepackage{fancyvrb}

\title[Properties of the stellar disks] %% give here short title %%
{Properties of the Milky Way stellar disks in the direction of the Draco dSph galaxy.}

\author[Anna S. \'{A}rnad\'{o}ttir, Sofia Feltzing \& Ingemar Lundstr\"{o}m]   %% give here short author list %%
{Anna S. \'{A}rnad\'{o}ttir$^1$, Sofia Feltzing$^1$ \and Ingemar Lundstr\"{o}m$^1$}

\affiliation{$^{1}$Lund Observatory, \\ Box 43,
SE-22100 Lund, Sweden \\ email: {\tt anna, sofia, ingemar @astro.lu.se} }

\pubyear{2008}
\volume{254}  
\pagerange{}
\jname{The Galaxy Disk in Cosmological Context}
\editors{J. Andersen, J. Bland-Hawthorn \& B. Nordstr\"{o}m, eds.}
\begin{document}

\maketitle

\begin{abstract}

We present the first results of a study where we determine the metallicity distribution 
function in the Galactic disks as a function of height above the Galactic plane.  
Observations in the Str\"{o}mgren photometric system enables us to identify the 
dwarf stars and derive metallicities for them.  
The resulting metallicity distribution functions at 0.5 and 2.0 kpc above the Galactic 
plane are significantly broader and more metal-rich than is anticipated from standard 
models such as the Besan\c{c}on model.  Our results can be explained by invoking 
a smaller scale height and larger local normalisation for the thick disk than is commonly 
used in the models. These results are compatible with recent determinations of the thick 
disk scale height based e.g. on SDSS data.

The age of the stellar populations as a function of height above the Galactic plane is also investigated by 
studying the turn-off colour and metallicity.  We tentatively find that at 2.0 kpc
above the Galactic plane there exist an intermediate age population.

\keywords{Galaxy: stellar content -- Galaxy: structure -- Galaxy: disk -- stars: distances }

\end{abstract}

\firstsection % if your document starts with a section,
             
\section{Introduction}

The Milky Way galaxy hosts two stellar disks.   The properties of the disks 
appear to differ in that the thick disk is older and more metal-poor than the
thin disk. Much of our knowledge of these properties for the disks
is acquired through the study of local stellar samples (e.g. 
\cite[Bensby et al. 2005]{Bensby05}, \cite[Fuhrmann 2008]{fuhrmann08}, 
\cite[Reddy et al. 2006]{reddy06}).

\begin{figure}[t]
% \vspace*{-2.0 cm}
\begin{center}
\scalebox{0.42}{ %
 \includegraphics{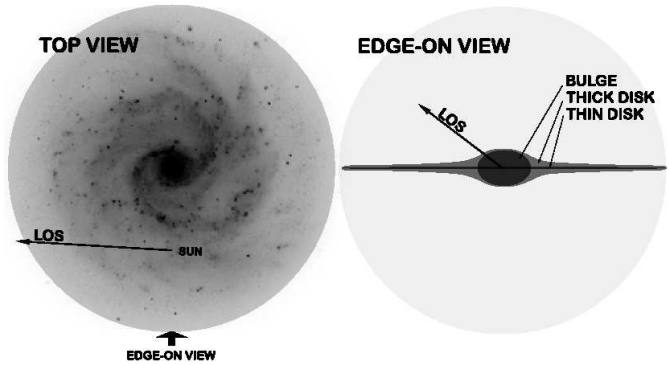} 
\vspace*{-1.0 cm}
}
\caption{Schematic figure showing the observed line-of-sight (LOS).} 
\label{figure1}
\end{center}
\end{figure}

The stellar sample studied here lies along the line-of-sight towards the 
Draco dwarf spheroidal galaxy, located at $l=86.4$ and $b=34.7$, Fig. \ref{figure1}.  
Along this line-of-sight the Milky Way disk stars are approximately at the same 
galactocentric distance as the sun.  
This means that we are fairly insensitive to any radial gradients that might exist in
the disk, and can assume that any gradients found are basically due to changes 
in the properties of the stellar populations in the $z$-direction.

This study uses the Str\"{o}mgren photometric system which is made up of four intermediate 
band filters ($uvby$) that are strategically located so that information can be obtained about  
the amount of line blanketing and the size of  the Balmer jump.  \cite[Str\"{o}mgren (1963)]{Stromgren63}
devised two indices, $m_1$ and $c_1$, that measures these effects.
While the $m_1$ index is metallicity sensitive, the $c_1$ index is 
sensitive to surface gravity and we use it to separate distant giant stars, belonging to the 
dwarf spheroidal galaxy, from Galactic dwarf stars and sub-giant stars, belonging to the thin disk,
thick disk, and halo, respectively.

\section{The data and parameters derived from the photometry}

The observations were obtained with the Isaac Newton Telescope on LaPalma using the Wide 
Field Camera.  Observations were taken both of the centre and the outskirts of the Draco 
dwarf spheroidal galaxy.  After carefully separating the background members of the dwarf
spheroidal galaxy
from the foreground dwarf stars and sub-giant stars we present here results based 
on the properties of 749 Galactic dwarf and sub-giant stars with $V=15-18.5$.  
%This magnitude cut was selected to ensure completeness.

For these stars we have determined both distances and metallicities 
based on Str\"{o}mgren photometry.  The metallicity calibration of
\cite[Ram{\'{\i}}rez \& Mel{\'e}ndez (2005)]{RM05} was used for stars with $(b-y) \le 0.8 $ and the
metallicity calibration of \cite[Olsen (1984)]{Olsen84} for stars with $0.8 < (b-y) \le 1.0$.
Of the available calibrations in the literature these best reproduce spectroscopic
iron abundances for a compiled set of 460 test stars (\cite[\'{A}rnad\'{o}ttir et al., in prep]{Arnadottir09}).
In this magnitude range the errors in the derived metallicities are below $\sim 0.3$ dex.

The line-of-sight toward the Draco dwarf spheroidal galaxy has been observed as 
part of the Sloan Digital Sky Survey (SDSS) so
$ugriz$ photometry is available for our sample of stars.  We have estimated the completeness
of our photometry
using the SDSS survey, and base our magnitude cut on that completeness estimate.
The distance modulii of individual stars were determined using $ugriz$ apparent magnitudes 
and absolute magnitudes estimated using the isochrones of  \cite[Girardi et al. (2004)]{Girardi2004}.  
For stars near the turnoff we calculated a $\log g$ using our own calibration 
(\cite[\'{A}rnad\'{o}ttir et al., in prep]{Arnadottir09}).  This $\log g$ was used to constrain the solution, 
i.e. we know if we were determining the distance to a main sequence dwarf or a sub-giant star.

The photometric $ugriz$ metallicity calibration of \cite[Karaali et al. (2005)]{Karaali05} was tested 
but it clearly is not able to derive the metallicities for the dwarf stars under consideration here 
and can therefore not be used for a larger investigation of the dwarf star properties in the disk 
based on $ugriz$ photometry.
Additionally, there is very little overlap between this study and the Sloan Extension for Galactic Understanding
and Exploration (SEGUE) survey.  Only 4 stars were found in common between the two.  

\begin{figure}[t]
% \vspace*{-2.0 cm}
\begin{center}
\scalebox{0.7}{ %
 \includegraphics{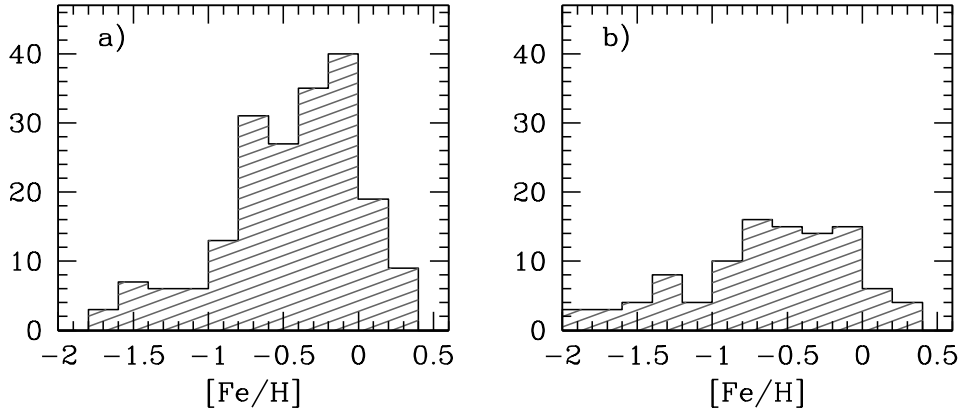} 
 \vspace*{-1.5 cm}
}
\caption{The metallicity distribution function for dwarf stars with $15<V<18.5$ is shown at heights
a) z=0.5 kpc ($\pm 0.2$ kpc) and b) z=2.0 kpc ($\pm 0.5$ kpc) above the Galactic plane
in the direction of $(l,b)=(86.4,34.7)$. Typical error in [Fe/H] is $<0.3$ dex.} 
\label{figure2}
\end{center}
\end{figure}

\section{Metallicity distribution  as a function of height above the Galactic plane and implications for the scale heights of the thin and the thick disk}

We present the metallicity distribution functions at 0.5 kpc and 2.0 kpc above the Galactic 
plane, Fig. \ref{figure2}. At both heights we see a broad distribution of metallicities.

The observed distributions are not well reproduced by
the Besan\c{c}on model of the Galaxy \cite[(Robin et al. 2003)]{Robin2003}.
The predicted metallicity distribution from the Besan\c{c}on model  
is narrower and at 0.5 kpc above the Galactic plane it is more peaked towards solar metallicities 
and the model predicts that 80\% of the stars are contributed from the thin disk 
and 20\% of the stars are from the thick disk.
At 2.0 kpc above the Galactic plane the predicted metallicity distribution from the Besan\c{c}on model is 
also narrower than the observed one, and peaks at lower metallicities, [Fe/H]$\sim -0.9$.  Here
the model predicts that 89\% of the stars are thick disk stars with only 7\% contribution from the
thin disk and 4\% from the halo.

In summary, our data show broader metallicity distribution functions than the Besan\c{c}on model
and our data also peaks at higher metallicities than the model does. 
Can we reconcile our data with a realistic model of the Milky Way disk system?  

\begin{figure}[t]
% \vspace*{-2.0 cm}
\begin{center}
\scalebox{0.58}{ %
\includegraphics{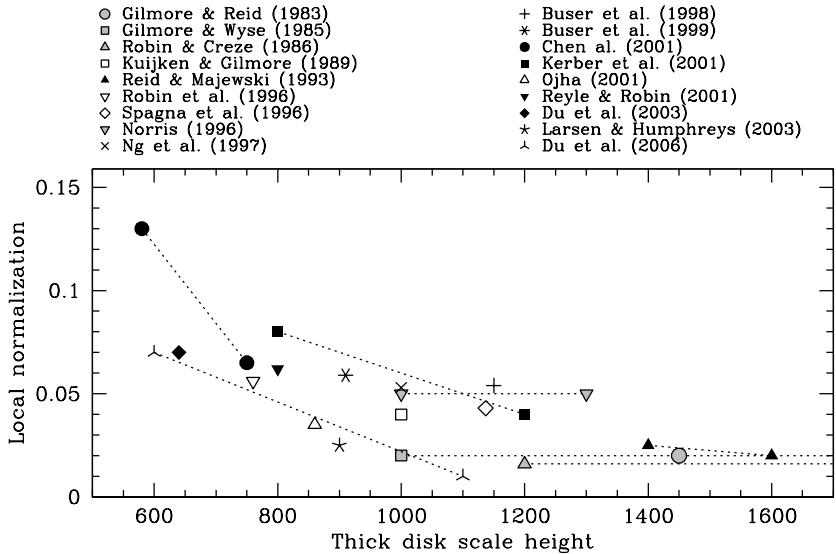} 
\vspace*{-1.0 cm}
}
\caption{The local normalisation is plotted against the thick disk scale height for a compiled list
of values from the literature.  References are given above the Figure.  When a study gives a range
of values, the extremes are plotted and connected with a dotted line.} 
\label{figure4}
\end{center}
\end{figure}

One thing that is not too well constrained in the literature are the scale heights of
the thin and the thick disks and the local normalisation of the stellar population densities.  
The thin disk scale height is determined to be between 220 pc and 330 pc (see i.e. 
\cite[Du et al. (2006)]{Du2006} and \cite[Chen et al. (2001)]{Chen2001}).
The thick disk scale height, varying between 580 pc and 1600 pc,
and its local normalisation, varying between 1\% and 13\%, are even more unconstrained.  
In Fig. \ref{figure4} we have compiled a collection of thick disk scale heights and local 
normalisations from the literature.  There is a large range of acceptable values.
The Besan\c{c}on model  \cite[(Robin et al. 2003)]{Robin2003} has a thick disk scale 
height of 800 pc and a local normalisation of 6.8\%.

Using a toy model where each stellar population, i.e. thin disk, thick disk and halo, is represented by 
a single age and single metallicity we varied the scale heights of the thin and the thick disk and the local 
normalisation.  This enabled us to study the contribution from stars ($15<V<18.5$) of each population 
to the metallicity distribution at 0.5 kpc and 2.0 kpc above the plane, and qualitatively compare  it to
the observed metallicity distributions shown in Fig. \ref{figure2}.  We tested several combinations
of thin and thick disk scale heights and local normalisation from the literature (Fig. \ref{figure4}).
The observed broad metallicity distributions, Fig. \ref{figure2}, suggest that the contribution from
the thin and the thick disk are much closer to 50-50\% than the model predicts.
At this stage our comparison can only be qualitative.  However, we believe that the results 
will indeed hold up also with the more detailed modelling that we are now undertaking.  
 
Our preliminary results from this simple modelling is that a thin disk scale height of $320-320$ pc,
a thick disk scale height of $580-640$ pc, and a local normalisation of $7-13$\% 
appear to explain our data.  This combination of scale heights is advocated 
by recent studies such as \cite[Chen et al. (2001)]{Chen2001}, which is based on star counts in
279 deg$^2$ in the SDSS survey, and \cite[Du et al. (2006)]{Du2006},
which are based on star counts in 21 fields of 1 deg$^2$ in the 
Beijing-Arizona-Taiwan-Connecticut (BATC) survey.
%Note that the thick disk scale height used by the Besan\c{c}on model
%is already on the lower side (Fig. \ref{figure4}), but still lower values are needed in order to 
%explain the observed metallicity distribution functions.

In order to try to assess whether the observed line-of-sight is representative for the 
Milky Way disk system we queried the SDSS for stars in eleven 
fields in the northern galactic hemisphere with $90<l<270$ and $35<b<90$ (including one field centred 
on the Draco dwarf spheroidal galaxy).  
We systematically compared colour magnitude diagrams and the colour distributions 
in these fields to the results from the Besan\c{c}on model.

We find that in each field and in the magnitude range $15<V<18.5$ the 
turn-off stars in the Besan\c{c}on model are
systematically bluer and overproduced as compared to the SDSS data.  
In this magnitude range we are, in the bluest part of the colour distribution, looking at the thick
disk turn-off stars at $\sim 1-3$ kpc above the Galactic plane.  The observed overproduction 
is in accordance with our previous indications that the true scale height of the thick 
disk is thinner than that used by the Besan\c{c}on model.  

Since this effect is observed systematically along several lines-of-sight 
we conclude that the line-of-sight towards the Draco dwarf spheroidal galaxy is not unique
and conclusions drawn here about the thick and thin disk scale heights from the metallicity 
distributions along that line-of-sight are of a more general nature and can be applied to the
Milky Way as a whole.

\section{Population age as a function of height above the Galactic plane}

\begin{figure}[t]
% \vspace*{-2.0 cm}
\begin{center}
\scalebox{0.75}{ 
 \includegraphics{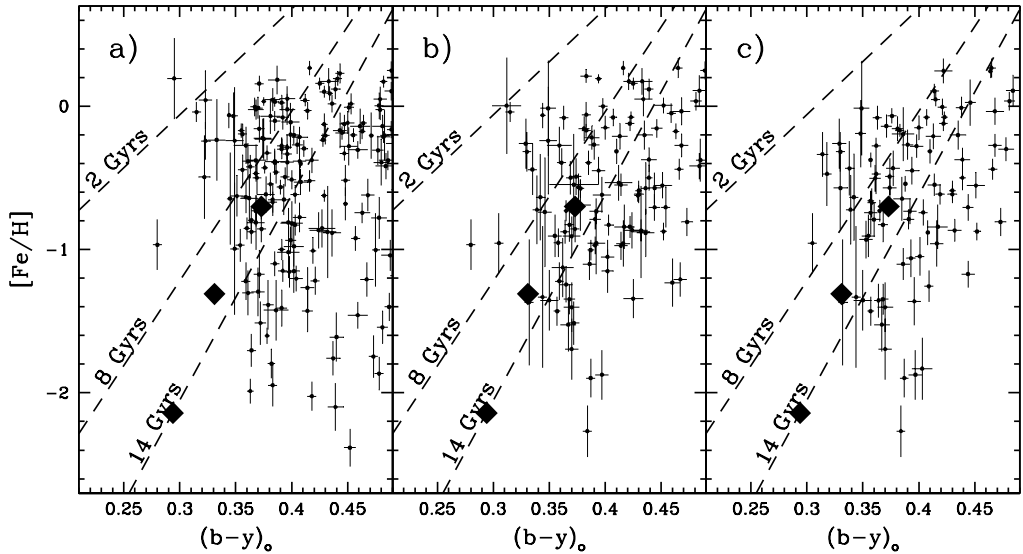} 
 \vspace*{-1.0 cm}
}
\caption{Metallicity is plotted against $(b-y)$ colour for stars with $\epsilon_{[Fe/H]}<0.5$ and $\log g > 3.8$ at three heights above the Galactic plane: a) $1.0 \pm 0.3$ kpc, b) $1.5 \pm 0.3$ kpc, and c) $2.0 \pm 0.5$ kpc. Dashed lines show turn-off colour and metallicity for three ages (2 Gyrs, 8 Gyrs and 14 Gyrs). The turn-off colours of the globular clusters M92, NGC288 and 47 Tuc are marked with diamonds.} 
\label{figure3}
\end{center} 
\end{figure}

We are able to study the turnoff colour as a function of height above the Galactic plane, 
enabling us to measure the age of the population and their metallicities as a function of height.
In Fig. \ref{figure3} we show the colour distribution of
dwarf stars at different heights above the Galactic plane.  Only dwarf stars are plotted.
The turnoff colour and metallicity are shown for 3 sets of ages, obtained using the
isochrones of  \cite[VandenBerg et al. (2006)]{VandenBerg2006}.  By determining where
the turn-off stars lie, one can read off the population age at that height above the
Galactic plane. For validation we mark three globular clusters, M92 ([Fe/H]$=-2.14$),
 NGC288 ([Fe/H]$=-1.31$), and 47 Tuc ([Fe/H]$=-0.70$) that can all assumed to be old, 
onto the plot \cite[(Calamida et al. 2007)]{calamida07}. 

Fig. \ref{figure3} shows an indication of an intermediate age population (younger than 8 Gyrs) 
present even at large heights above the galactic Disk in the direction towards 
the Draco dwarf spheroidal galaxy (the direction our observations were taken in).

\section{Summary}

Using Str\"{o}mgren photometry to derive the stellar parameters of dwarf stars
in the Milky Way along the line-of-sight towards the Draco dwarf spheroidal 
galaxy we have studied the properties of the stellar populations in the Milky 
Way disks and find that

\begin{itemize}
\item The metallicity distribution functions at 0.5 kpc and 2.0 kpc above the Galactic plane are 
broader than predicted by the Besan\c{c}on model and the observed metallicity gradient as
a function of height above the Galactic plane is not as steep as predicted.
\item Preliminary investigations, using a simple toy model, indicates that our observations favour 
models with smaller thick disk scale heights and higher local normalisation such as found by 
e.g. \cite[Chen et al. (2001)]{Chen2001}.
\item By studying the turn-off colour and metallicity as a function of height above the 
Galactic plane, we find an indication of an intermediate age population located 
2 kpc above the Galactic plane.
\end{itemize} 

Further modelling of the data is being undertaken, but we believe that these preliminary results
will hold.  

\begin{acknowledgments}

The Royal Physiographic Society in Lund supported AAs attendance at this symposium
with a travel grant for young researchers.

 SF is a Royal Swedish Academy of Sciences Research Fellow supported by a 
grant from the Knut and Alice  Wallenberg Foundation. 

\end{acknowledgments}

%\begin{discussion}
%\discuss{Someone}{Some comment.}
%\end{discussion}

\end{document}